# Quantised Vortices in an Exciton-Polariton Fluid

K. G. Lagoudakis[1], M. Wouters[2], M. Richard[1], A. Baas[1], I. Carusotto[4], R. André[3], Le Si Dang[3], B. Deveaud-Pledran[1]

[1] IPEQ, Ecole Polytechnique Fédérale de Lausanne(EPFL), Station 3, 1015 Lausanne, Switzerland.

[2] ITP, Ecole Polytechnique Fédérale de Lausanne(EPFL), Station 3, 1015 Lausanne, Switzerland. Previously :TFVS, Universiteit Antwerpen, Universiteitsplein 1, 2610 Antwerpen, Belgium.

[3] Institut Néel, CNRS, 25 Avenue des Martyrs, 38042 Grenoble, France.

[4] INFM-CNR BEC and Dipartimento di Fisica, Universita di Trento, via Sommarive 14 38050 Povo (Trento) ITALY

**One of the most striking quantum effects in a low temperature interacting Bose gas is superfluidity. First observed in liquid $^4$He, this phenomenon has been intensively studied in a variety of systems for its amazing features such as the persistence of superflows and the quantization of the angular momentum of vortices[1]. The achievement of Bose-Einstein condensation (BEC) in dilute atomic gases[2] provided an exceptional opportunity to observe and study superfluidity in an extremely clean and controlled environment. In the solid state, Bose-Einstein condensation of exciton polaritons has now been reported[3,4,5,6] several times. Polaritons are strongly interacting light-matter quasi-particles, naturally occurring in semiconductor microcavities in the strong coupling regime and constitute a very interesting example of composite bosons. Even though pioneering experiments have recently addressed the propagation of a fluid of coherent polaritons[7], still no conclusive**

**evidence is yet available of its superfluid nature. In the present Letter, we report the observation of spontaneous formation of pinned quantised vortices in the Bose-condensed phase of a polariton fluid by means of phase and amplitude imaging. Theoretical insight into the possible origin of such vortices is presented in terms of a generalised Gross-Pitaevskii equation. The implications of our observations concerning the superfluid nature of the non-equilibrium polariton fluid are finally discussed.**

Vortices in superfluids carry quantised phase winding and circulation of the superfluid particles around their core. By definition, vortices are characterised by i) a rotation of the phase around the vortex by an integer multiple of $2\pi$, commonly known as the topological charge of the vortex and ii) the vanishing of the superfluid population at their core. Due to their major importance for the understanding of superfluidity, they have been intensively studied theoretically[8] and experimentally[9,10,11] in disorder free, stirred three dimensional BECs of dilute atomic gases and in quasi two-dimensional BECs where they spontaneously emerge from thermal fluctuations[12,13] and are strictly related to the Berezinskii-Kosterlitz-Thouless phase transition[14]. In the present work, we observed the spontaneous appearance of pinned singly quantised vortices as an intrinsic feature of non-equilibrium polariton BECs in the presence of disorder. Here BEC of polaritons is created in the same planar CdTe microcavity that was used in our previous studies[3,15,16]. The creation of the condensate was performed by means of non-resonant continuous wave optical excitation, the intensity of which is used here to drive the polaritons through the phase transition. Its steady state is characterised by the balance of an incoming and outgoing flow of polaritons. As a consequence, in contrast to the atomic BEC, the polariton condensate is intrinsically in a non-equilibrium situation. Moreover, the polariton condensate is created in a disordered environment, causing a spatial modulation of the condensate density, yet preserving its spatial long range order[3,16].

Access to the phase and amplitude of the polariton fluid order parameter is achieved via the observed polariton luminescence. Thanks to the photonic component of the cavity polariton, all the statistical properties of the intracavity polariton field are indeed mapped into the corresponding ones of the emitted luminescence[17]. The straightforwardness with which the order parameter characteristics are reached is a major advantage of microcavities with respect to all other quantum condensed systems. In our previous experiments[3,15,16] we used cross correlation techniques to probe the spatially resolved phase correlations of the polaritonic condensate. This method is similar to the one used in atomic BECs where a matter-wave interferogram is obtained by combining the fields originating from two expanding condensates[18]. Here, we use the same technique to probe the existence of vortices and to identify the characteristic $2\pi$ phase rotation. For this purpose, the luminescence of the condensate at the position of the vortex is overlapped with a different region of the condensate luminescence which does not contain any vortex and features a spatially homogeneous phase. This results in an interferogram which carries the phase of the macroscopic wave function in the vortex region. The $2\pi$ phase rotation carried by the vortex, gives rise to a forklike dislocation in the interference pattern as shown in figure 1a,b. For atomic BECs in the presence of vortices, similar dislocations have been observed experimentally[11] and modelled theoretically[19,20].

By moving the excitation spot on the sample, we systematically observe vortices with unit topological charge as evidenced by the single $2\pi$ phase winding around their core as shown in figure 1c,d. Note that the observed phase winding extends up to a region far from the vortex core, which makes the $2\pi$ rotation of the phase unambiguous. The precise phase value measured along closed circulation loops of various radii around the vortex are plotted in Fig.1d. In a homogeneous condensate the phase would be a linear function of the azimuthal angle[8] whereas here the polariton vortex is formed in a disordered medium which distorts this behaviour. In order to be confident that the observed dislocation is due to the phase singularity of a polariton vortex, we performed the same interferometric measurement several times, overlapping the vortex with different areas of the condensate (without a vortex). We additionally changed the orientation of the interference pattern. Whatever the direction chosen, we were always able to distinguish the forklike dislocation precisely centred at the position of the vortex. This property rules out the interpretation of the forklike interference pattern in terms of trivial optical effects like point-like or line-like structural defects of the microcavity affecting the polariton emission wavefront. This property is illustrated in figure 1b: the vortex is the same as in figure 1a but here it is overlapped with a different region of the condensate and with a different overlap angle (rotated fringes). In this way we demonstrate unambiguously the existence of the phase winding which is the main characteristic property of quantised vortices.

As we already mentioned, another key characteristic of the vortices is the absence of superfluid density in their core. As the observed luminescence comes not only from the condensed state which contains the vortex but also from non-condensed hot polaritons (our real space images are the result of the integration from -30° to 30° by the NA=0.5 of our microscope objective), we had to isolate the condensate polaritons by spectrally resolving the real space luminescence image. The spatial distribution of polaritons at the energy of the condensate is shown in figure 2a. The vortex position as defined by the singularity in the spatial phase profile does indeed correspond to a local minimum of the condensate density, as seen in figure 2b. Instead of a complete vanishing of the population, we only observed a slight reduction of the local density with respect to the surrounding population. Knowing that our optical resolution is limited by diffraction to about 1 μm, this measurement is consistent with a strong reduction in density at the vortex core over a diameter smaller than 1μm.

A peculiarity of the vortices in the non-equilibrium polariton condensate is the fact that they appear spontaneously, without any stirring of the condensate. Because of the two-dimensional nature of the polariton condensate, one might naively think of vortices being spontaneously formed as a consequence of thermal fluctuations, as recently observed in quasi two-dimensional ultracold atomic BECs[12]. This mechanism is however contradicted by the experimental data as the interference patterns are averaged over many runs of the condensate formation. Although a scenario in which the spontaneously formed vortex is trapped by the disorder could account for a well defined vortex position, its random sign would lead to a decrease of the fringe contrast in the neighbourhood of the vortex, contrary to what we observe here. An additional argument against the spontaneous proliferation of thermal vortices in the present polariton condensate is the fact that the spatial coherence function (not shown) exhibits a plateau for large distances, rather than a fast exponential decay.

The mechanism that is responsible for the spontaneous appearance of the vortex has therefore to be deterministic. The emergence of deterministic flow in polariton non-equilibrium condensates has been previously attributed to the combined effect of the continuous pumping and inhomogeneity of the system[21,25] : in marked contrast with equilibrium condensates whose ground state is always flowless, the presence of pumping and losses makes the condensate polaritons constantly flow down the hills of the disorder potential landscape. Depending on the details of the disorder potential, the non-equilibrium flow pattern may show vortex singularities. This simple physical picture of the emergence of vortices is supported by numerical simulations based on the mean field model recently developed for non-equilibrium condensates in reference 22. An example is shown in figure 3. A vortex singularity is clearly seen in the density and phase profile. The arrows, which represent the local wave vector $k = \nabla \phi$, wind around the vortex core. In addition, the vortex core appears to be of the order of 1 μm, which corresponds approximately to both the characteristic length scale associated to the damping rate $l_\gamma = (\hbar/\gamma m)^{1/2}$ and the healing length $\xi = (\hbar/g|\psi|^2 m)^{1/2}$ obtained from the values used in the theoretical simulations.

It would be obviously tempting to consider the observation of persistent quantised vortices in the polariton condensate as a proof of its superfluidity. Unfortunately, no clear cut criteria for superfluidity in a non-equilibrium system such as ours are yet available. Hence, before using the term superfluidity, a better understanding of the very concept of superfluidity in a non-equilibrium context [26] is needed. It is however safe to conclude that our polariton fluid shares with conventional superfluids such as liquid $^4$He and Bose-Einstein condensed ultracold atomic gases the crucial property of having a quantised vorticity.

In conclusion, we have reported on the experimental observation of singly quantised vortices in an exciton polariton condensate. A clear $2\pi$ phase shift phase singularity has been extracted through interferometric techniques. This quantised phase variation is accompanied by a clear reduction of the polariton fluid density at the vortex location. A theoretical description of the polariton condensate within a mean field model suggests that the vortices arise from the interplay between the disordered and driven-dissipative nature of the polariton condensate.

METHODS

**Experimental realisation.** The experimental setup is an improved version of the one used in our previous studies[3]. In short, the sample is the same CdTe microcavity cooled down to 4.2 °K by means of a cold finger liquid helium flow cryostat. We excited our system in a non-resonant quasi-continuous wave manner in order to avoid heating of the sample. The excitation beam was set at the wavelength of the first minimum of the reflection of the sample in order to couple efficiently light in the intracavity. The luminescence of the condensate was observed with a high numerical aperture (NA=0.5) microscope objective allowing the achievement of diffraction limited 2D imaging, in conjunction with an interferometric system in order to be able to acquire the luminescence interferogram. For this purpose the luminescence was sent in an actively stabilised Michelson interferometer with a mirror-retroreflector configuration (for details see reference 3) and the real space interferogram of the luminescence was imaged on a high resolution CCD.

The procedure for the phase extraction of the interferogram is analogous to the one in reference 23 were the phase is extracted from the time-frequency domain. Here the transform is in two dimensions and for real space - k space. The presence of fringes in the luminescence interferogram is of crucial importance because in this way the two conjugate parts of the Fourier Transform of the interferogram become separated in the Fourier space. The phase of the interferogram is carried in each of the parts, and by isolating one of them we were able to retrieve the phase information of the interferogram. The retrieved spatial phase profile was then compared to the phase profile of an interferogram without phase singularities but of the same fringe spacing. This difference reveals any phase anomalies or singularities carried in the interferogram.

Spectrally resolved imaging was performed on the same position of the sample by imaging the real space luminescence on the entrance slits of the ~10μeV resolution double monochromator. The spectrally resolved real space is then reconstructed from a sufficient number (~120) of spectrally resolved real space lines. In that way we were able to extract information on the spatial density profile at the energy of the condensate.

**Theoretical description.** The theoretical model that is used to explain the spontaneous appearance of deterministic vortices was introduced in Reference 22; a related method was proposed and applied to polariton condensates containing a vortex in Reference 21. Our mean field model is based on a generalised Gross-Pitaevskii equation for the low-

energy polariton states

$$i\frac{\partial}{\partial t}\psi(\vec{r},t) = \left\{-\frac{\hbar\nabla^2}{2m} + V_d(\vec{r}) - \frac{i}{2}\left[\gamma_c - R(n_R(\vec{r},t))\right] + g|\psi(\vec{r},t)|^2 + g_R n_R(\vec{r},t)\right\}\psi(\vec{r},t)$$

that takes into account the dissipation of polaritons at a rate $\gamma_c$ and their replenishing by stimulated scattering from the exciton reservoir at a rate $R(n_R)$ that is a function of the exciton reservoir density $n_R$. The photonic disorder is modelled by the potential term $V_d$. It is extracted from spatially and energetically resolved photoluminescence measurements far below the condensation threshold. In a first approximation, the lowest emission energy in the luminescence spectrum at a given position corresponds to the local potential energy. Elastic interactions between the condensate polaritons among themselves cause a blue shift of the condensate energy by $g|\psi|^2$ and interactions of the condensate polaritons with the reservoir excitons by an amount $g_R n_R$. The equation for the condensate field has to be coupled to a motion equation for the reservoir density. As a simple model, we adopt a rate equation of the form

$$\frac{\partial}{\partial t}n_R(\vec{r},t) = P(\vec{r},t) - \gamma_R n_R(\vec{r},t) - R(n_R(\vec{r},t))|\psi(\vec{r},t)|^2$$

that describes the balance between the pumping rate $P$, the loss from the reservoir and stimulated scattering into the condensate states. This model was shown to recover the elementary excitation spectrum of nonequilibrium condensates predicted by the Keldysh Green function method[24], as well as the main experimental spatial and spectral features of polariton condensates pumped with a finite excitation spot[25]. The parameters appearing in the equations are phenomenological and should in principle be determined from the experiment. Unfortunately, not sufficient experimental data are available to univocally determine them. The parameters that were used in the simulations were taken: $m\hbar^{-2}$ = 1.7meV$^{-1}$μm$^{-2}$, $\hbar\gamma$=1meV, $\hbar\gamma_R$=10meV, $\hbar g$=0.04meVμm$^2$, $\hbar R(n_R)$=(0.1meVμm$^2$)$n_R$, $\hbar g_R$=0.05meVμm$^2$, $\hbar P$=60meVμm$^{-2}$. Fortunately, the possibility for the appearance of vortices is not very sensitive to the precise values of the model parameters.

We thank D. Sarchi, V. Savona, J. Tempere and J. Devreese for fruitful discussions. The work was supported by the Swiss National Research Foundation through the "NCCR Quantum Photonics".

Correspondence and requests for materials should be addressed to K.G.L (konstantinos.lagoudakis@epfl.ch) or to M.W. (michiel.wouters@ua.ac.be)

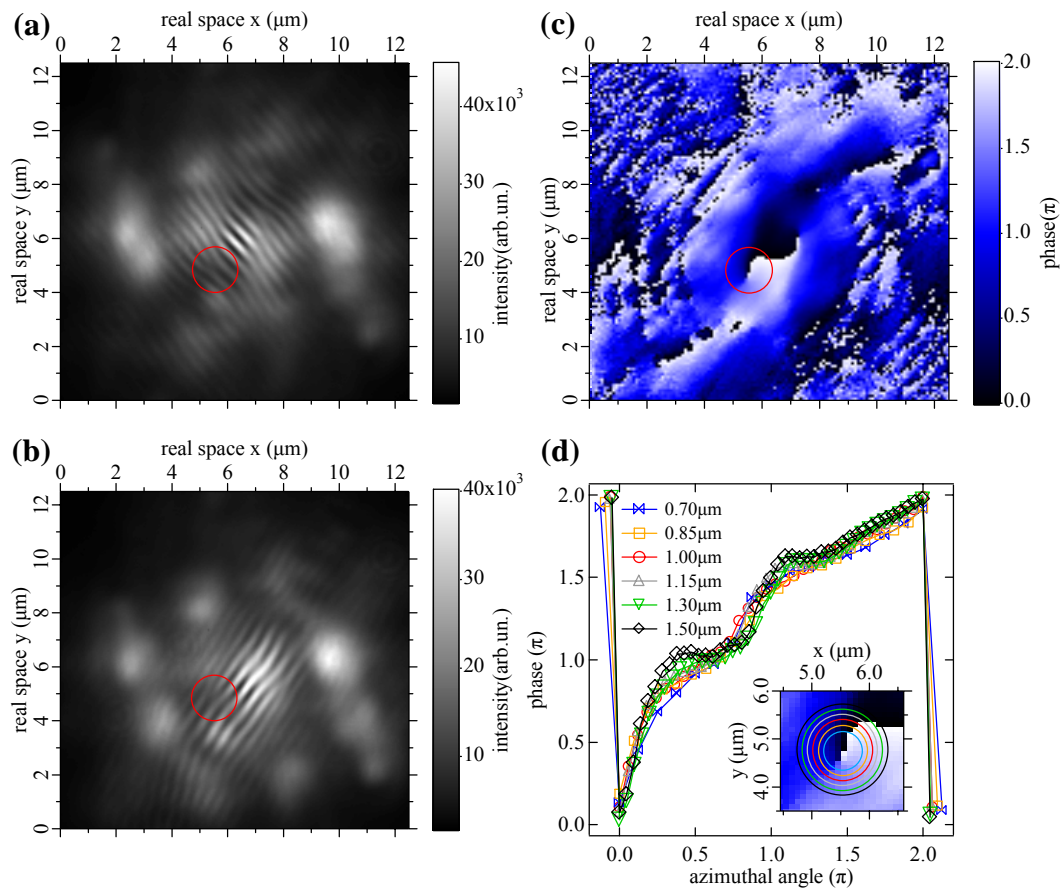

**Figure 1 Interferogram and extracted phase. a**, Interferogram with vortex: in the red circle one can see the forklike dislocation. **b**, Interferogram carrying the same information but this time the vortex is overlapped with a different region of the condensate and for different fringe orientation. The vortex appears at the same real space coordinates as before. **c**, Real space phase profile calculated from interferogram of (a). The red circle encloses the vortex (same real space area as on (a),(b)). **d**, Phase as a function of the azimuthal angle for a range of different radii as shown in the indent of figure (d) (zoom of (c)). Note that the data are repeated before and after the azimuthal angles 0 and 2π to better illustrate the 2π shift.

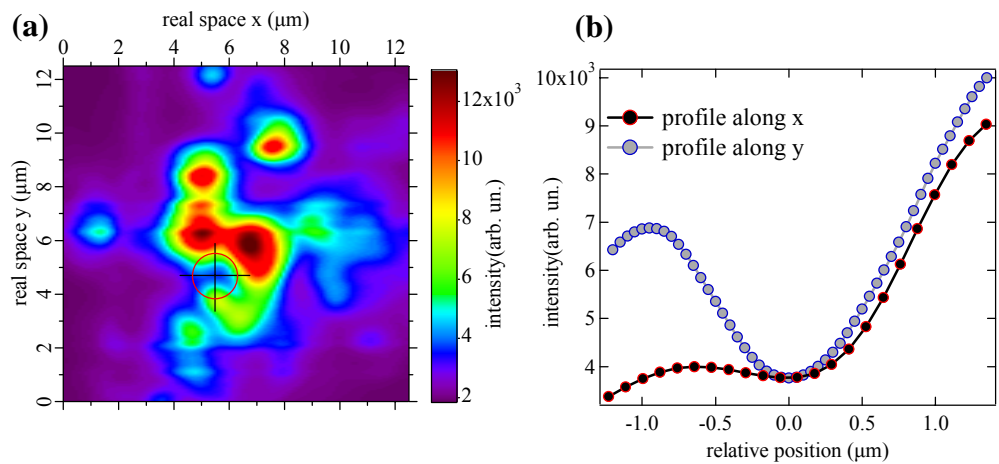

**Figure 2 Real space polariton population at the vortex region. a**, Two dimensional population at the energy of the condensate, over the hole excitation spot showing the location of the vortex in real space (centre of red circle). The integration in energies was done within the linewidth of the condensate (γ=650µeV). **b**, Population along the two arms of the black cross (black lines on (a)). The vortex is located at the position were the two lines cross each other, corresponding to a local minimum of the population.

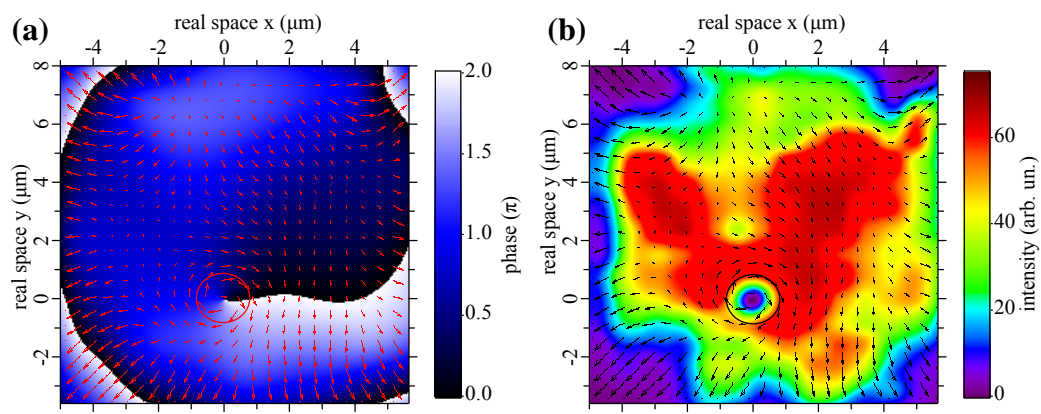

**Figure 3 Phase and density distribution from the mean field theory. a**, Theoretical phase profile from a simulation with the generalised Gross-Pitaevskii equation in the presence of a disorder potential. The arrows

representing the local wave vector $k=\nabla\phi$ wind around the branch point singularity in the phase at the vortex position. **b,** The real space density profile drops to zero at the centre of the vortex core.